\documentclass[aps, prd, twocolumn, superscriptaddress, preprintnumbers, longbibliography,nofootinbib]{revtex4-1}

\usepackage[top=2.8cm, bottom=2.8cm, left=2cm, right=2cm]{geometry}
\usepackage[utf8]{inputenc} 
\usepackage[T1]{fontenc}
\usepackage{amsmath, amssymb, lmodern} 
\usepackage[caption = false]{subfig}
\usepackage{xcolor}
\usepackage{graphicx}
\usepackage{natbib}
\usepackage{booktabs}
\usepackage{color}
\usepackage{hyperref}
\usepackage{tabularray}
\hypersetup{colorlinks, citecolor = blue, filecolor = blue, linkcolor = blue, urlcolor = blue}

\usepackage[normalem]{ulem}
\newcommand{\redsout}{\bgroup\markoverwith{\textcolor{red}{\rule[0.5ex]{2pt}{0.8pt}}}\ULon}
\definecolor{darkgreen}{rgb}{0, 0.6, 0}
\begin{document}

\preprint{IFT-UAM/CSIC-23-87}

\title{Machine learning unveils the linear matter power spectrum of modified gravity}

%%%%%%%%%%% AUTHORS INFO %%%%%%%%%%%%%%%%%%%%%%
\author{J. Bayron Orjuela-Quintana}
\email{john.orjuela@correounivalle.edu.co}
\affiliation{Departamento  de  F\'isica,  Universidad  del Valle, Ciudad  Universitaria Mel\'endez,  Santiago de Cali  760032,  Colombia}

\author{Savvas Nesseris}
\email{savvas.nesseris@csic.es}
\affiliation{Instituto  de  F\'isica  Te\'orica  UAM-CSIC,  Universidad  Auton\'oma  de  Madrid,  Cantoblanco,  28049  Madrid,  Spain}

\author{Domenico Sapone}
\email{domenico.sapone@uchile.cl}
\affiliation{Cosmology and Theoretical Astrophysics group, Departamento de F\'isica, FCFM, Universidad de Chile, Blanco Encalada 2008, Santiago, Chile}
%%%%%%%%%%% AUTHORS INFO %%%%%%%%%%%%%%%%%%%%%%

%%%%%%%%%%%%%%%%%%
\begin{abstract}
%%%%%%%%%%%%%%%%%%

The matter power spectrum $P(k)$ is one of the main quantities connecting observational and theoretical cosmology. Although for a fixed redshift this can be numerically computed very efficiently by Boltzmann solvers, an analytical description is always desirable. However, accurate fitting functions for $P(k)$ are only available for the concordance model. Taking into account that forthcoming surveys will further constrain the parameter space of cosmological models, it is also of interest to have analytical formulations for $P(k)$ when alternative models are considered. Here, we use the genetic algorithms, a machine learning technique, to find a parametric function for $P(k)$ considering several possible effects imprinted by modifications of gravity. Our expression for the $P(k)$ of modified gravity shows a mean accuracy of around $1$-$2\%$ when compared with numerical data obtained via modified versions of the Boltzmann solver \texttt{CLASS}, and thus it represents a competitive formulation given the target accuracy of forthcoming surveys.

%%%%%%%%%%%%%%%%%%
\end{abstract}
%%%%%%%%%%%%%%%%%%

\maketitle

%%%%%%%%%%%%%%%%%%%%%%%%%%%
\section{Introduction} 
\label{Sec: Introduction}
%%%%%%%%%%%%%%%%%%%%%%%%%%%

Since its designation as the standard cosmological model around two decades ago \cite{Peebles2020}, the model assuming the existence of the cosmological constant $\Lambda$ and a Cold Dark Matter (CDM) component, $\Lambda$CDM, has passed many stringent observational tests \cite{Planck:2018vyg, DES:2017qwj, Abbott:2018xao, DES:2021wwk, Riess:1998cb, perlmutter:1998np, SupernovaSearchTeam:2004lze, percival_2010, Blake_2011, Aubourg:2014yra, deBernardis:2000sbo, Jaffe:2003it, Planck:2018jri, Planck:2019evm, DES:2018ekb}. However, intriguing discrepancies have arisen in recent years \cite{Perivolaropoulos:2021jda, Abdalla:2022yfr}. Arguably, the most representative among these discrepancies is the $H_0$ tension whose significance is now around $5\sigma$ \cite{Wong:2019kwg, Riess:2020fzl, Riess:2021jrx}. Therefore, the search for a sound theoretical candidate is still plausible, highlighting Modified Gravity (MG) theories as one the most prominent alternatives\footnote{We want to stress that Modified Gravity theories and dark energy models can be treated on equal footing under some approaches \cite{Arjona:2018jhh, Arjona:2019rfn, Cardona:2022lcz, Nesseris:2022hhc, Bloomfield:2012ff, Gubitosi:2012hu, Frusciante:2019xia, Kunz:2006ca}, and thus a clear distinction between them is not necessary.} \cite{diValentino:2021izs, CANTATA:2021ktz}.

The clustering of matter, due to gravitational instabilities, forming the structure of our Universe constitutes a key physical process in the expansion history. This process offers one of the main links between observations and theoretical models through the so-called matter power spectrum $P(k, z)$, which in general is a function of the wavenumber $k$ and the redshift $z$ \cite{SDSS:2003tbn, BOSS:2016wmc, dodelson2020}. For a given redshift, $P(k)$ can be numerically computed using Boltzmann solvers, like the codes \texttt{CAMB} \cite{Lewis:1999bs} or \texttt{CLASS} \cite{Blas_2011}. Nonetheless, it is desirable to have an analytical formulation of $P(k)$ in some cases; for instance, when computing costs are demanding and high accuracy is not required. When $\Lambda$CDM is the assumed cosmological model, there are two well-known fitting formulae to determine the transfer function involved in the computation of $P(k)$. These are the Bardeen-Bond-Kaiser-Szalay (BBKS) formula \cite{Bardeen:1985tr} and the Eisenstein-Hu (EH) formulae \cite{Eisenstein:1997ik, Eisenstein:1997jh}, which are widely used in the literature \cite{Dvornik:2022xap, Valles-Perez:2023cgf, Campagne:2023ter, Petter:2023imh, Euclid:2023uha, Prole:2023iyh, Variu:2022peh, Hutsi:2022fzw}. However, up to the best of our knowledge, there are no similar formulations considering alternative models. 

In Ref.~\cite{Orjuela-Quintana:2022nn} two of the authors here used a machine learning technique known as the Genetic Algorithms (GAs) to find an analytical expression for the matter power spectrum under the $\Lambda$CDM framework and neglecting the representative wiggles of $P(k)$ due to the BAOs (Baryon Acoustic Oscillations). This new formula shows to be as accurate as its predecessors while being considerably simpler; and thus, easier to implement in other numerical computations where the knowledge of the de-wiggled $P(k)$ is required. Here, we want to extend that work by considering some possible effects on $P(k)$ introduced by modifications of gravity. Briefly, using the GAs, we find a parametric function for the de-wiggled $P_\text{MG}(k)$ whose performance is tested by comparing against numerical data obtained from the popular Boltzmann solver \texttt{CLASS}. In particular, we use the branches \texttt{hi\_class} \cite{Zumalacarregui:2016pph, Bellini:2019syt} and \texttt{mgclass} \cite{Baker:2015bva, Sakr:2021ylx}. Both allow to numerically compute $P(k)$ for MG theories. The former considers the whole Horndeski theories, while the later uses the slip parameters framework.

The paper is organized as follows. In the next section we give some generalities about the GAs. Then, in Sec.~\ref{Sec: The Linear Matter Power Spectrum of Modified Gravity}, we describe the matter power spectrum $P(k)$ in general and the possible effects due to modifications of gravity we take into account. In Sec.~\ref{Sec: Results}, we present the parametric function for the $P(k)$ of MG theories, discuss its applicability and its accuracy when compared against numerical data. Finally, in Sec.~\ref{Sec: Conclusions}, we give our concluding remarks.

%%%%%%%%%%%%%%%%%%%%%%%%%%%%%%%%%
\section{Genetic Algorithms}
\label{Sec: Genetic Algorithm}
%%%%%%%%%%%%%%%%%%%%%%%%%%%%%%%%%

The Genetic Algorithms (GAs) are an unsupervised machine learning technique in which the extraction of features and patterns from a given dataset is performed by mimicking biological evolution processes \cite{koza1992genetic}. In a nutshell, a random initial population to fit the data is created. Then, GAs improve their description of the data by implementing a natural selection scheme based on reproduction and survival of the fittest individuals. This process, that is stochastic in essence, is carried out for as many generations as desired or needed to achieve the expected accuracy of the fitting. 

The working principle of the GAs makes them an excellent tool as a symbolic regression approach. In a few words, symbolic regression consists in discovering a fitting function that accurately matches a given dataset. Now we succinctly describe how to GAs can be used in symbolic regression problems. Firstly, the initial individuals, corresponding to random mathematical expressions, are created. Then, the fitness of each individual in this first generation is measured, which can be done by using a simple mean squared error, for example. A portion of the fittest individuals are selected from a tournament scheme. The winners of this tournament survive to the next generation to create offspring through the so-called genetic operators, usually corresponding to crossover between individuals and mutation. In the subsequent generations, consisting of the most capable individuals from the previous generation and their offspring, the evolutionary process is repeated until an individual achieves the desired accuracy for data description. See Refs.~\cite{koza1992genetic, Nesseris:2012tt} for more details about GAs.

Finally, we point out that GAs have been used in several branches of physics, such as particle physics \cite{Allanach:2004my, Akrami:2009hp, Abel:2018ekz}, astrophysics \cite{Luo:2019qbk}, and astronomy \cite{ho:2019zap}. Particularly in cosmology, they have been used as a symbolic regression tool \cite{Orjuela-Quintana:2022nn, Aizpuru:2021vhd}, but also used to perform null tests of cosmological datasets \cite{Nesseris:2010ep, Arjona:2021hmg}, constraints on cosmological quantities \cite{Bogdanos:2009ib, Arjona:2019fwb, Arjona:2020kco, arjona:2021mzf, Alestas:2022gcg}, and forecasts for upcoming surveys \cite{Arjona:2020axn, EUCLID:2020syl, Euclid:2021frk}. There are also other approaches to symbolic regression based on different methodologies as described in e.g. Refs.~\cite{schmidt:2019doi, brunton:2016dac, Udrescu:2019mnk, Liu:2021azq, cranmer:2020wew, Bartlett:2022kyi}.

\begin{figure*}
\centering
\begin{minipage}[b]{.45\textwidth}
\includegraphics[width=\textwidth]{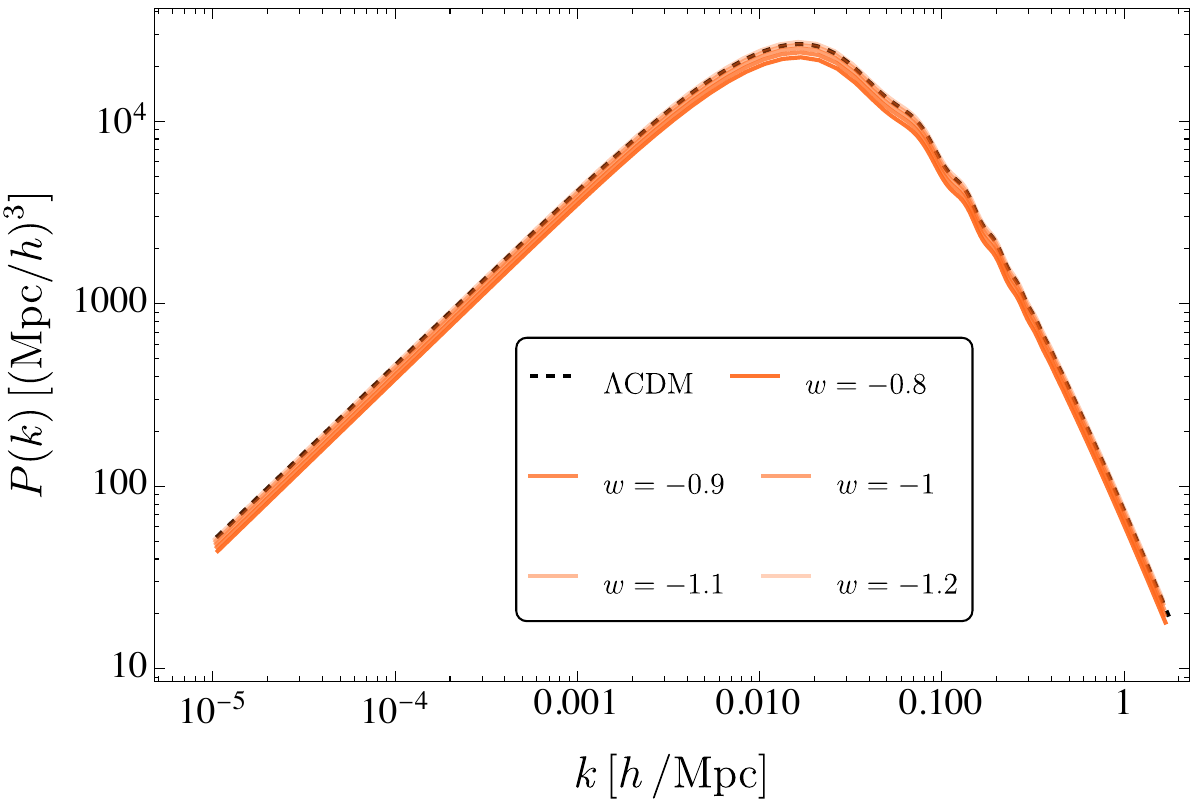}
\end{minipage}
\qquad
\begin{minipage}[b]{.45\textwidth}
\includegraphics[width=\textwidth]{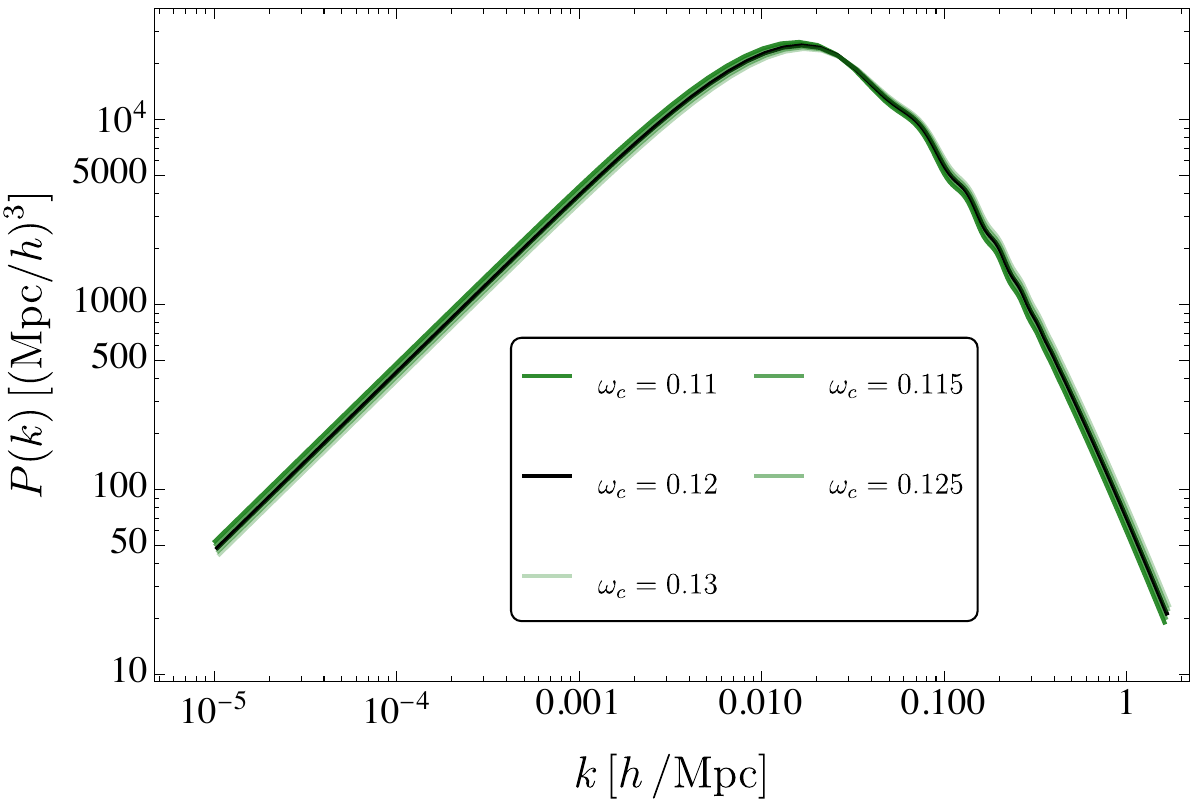}
\end{minipage}
\par\bigskip % force a bit of vertical whitespace
\begin{minipage}[b]{.45\textwidth}
\includegraphics[width=\textwidth]{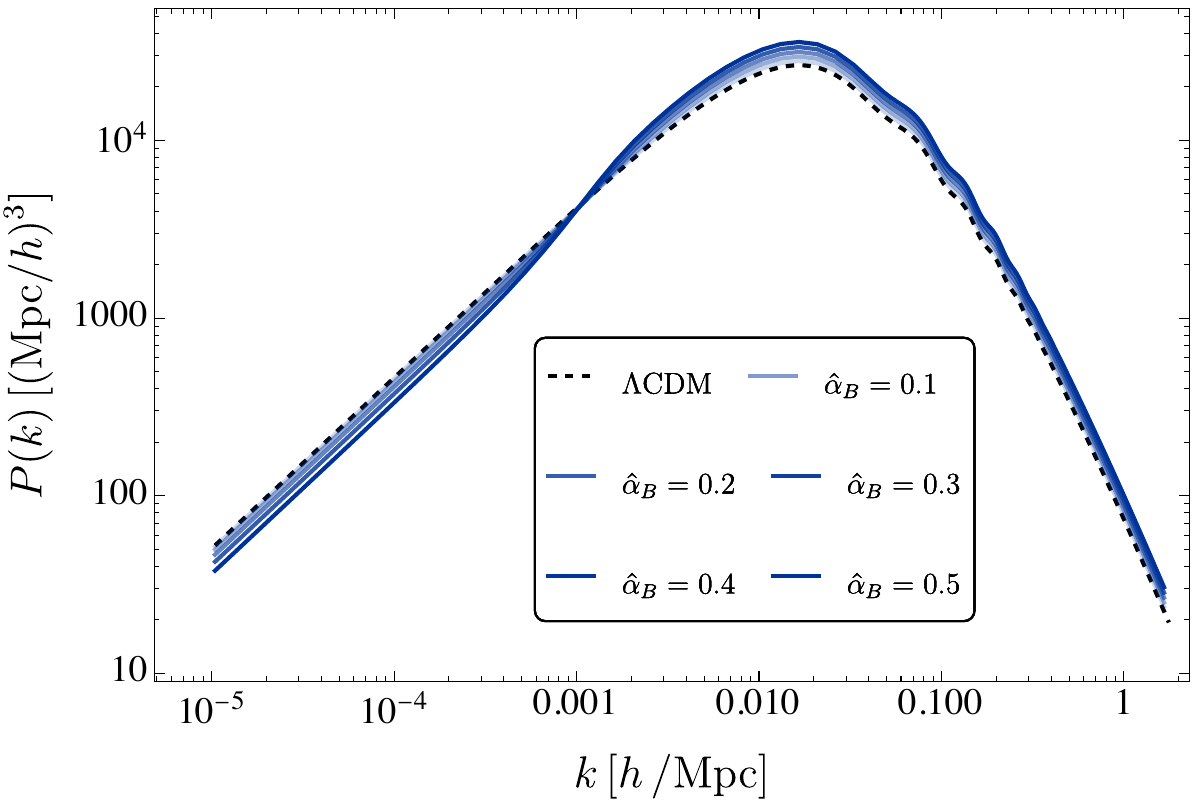}
\end{minipage} 
\qquad
\begin{minipage}[b]{.45\textwidth}
\includegraphics[width=\textwidth]{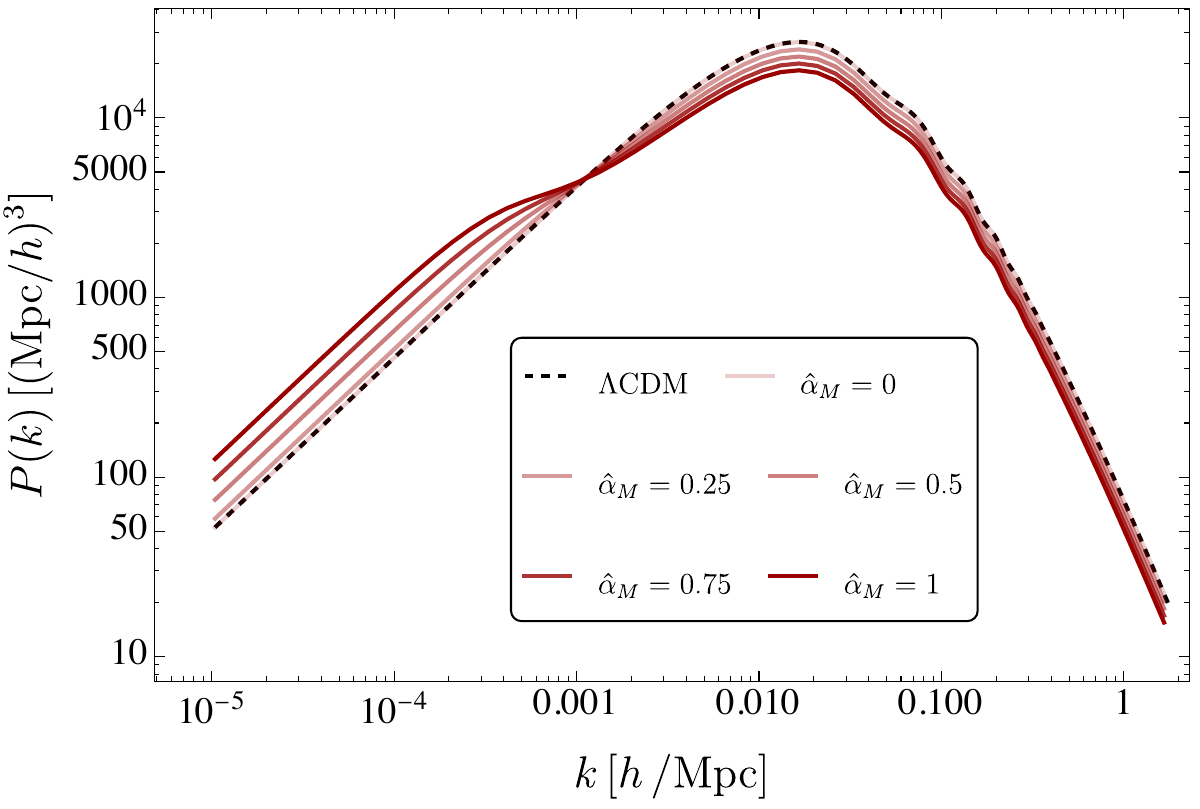}
\end{minipage}
\caption{Typical effects of modified gravity on the theoretical matter power spectrum. In the upper panels, we consider a $w$CDM model varying the dark energy equation of state $w$ (left), and the reduced CDM density parameter $\omega_c$ (right). In the lower panels, we consider the \texttt{proto\_scale} models in \texttt{hi\_class}, varying the braiding $\hat{\alpha}_B$ (left) and the mass running $\hat{\alpha}_M$ (right) parameters. In the former case, the matter power spectrum is suppressed at very large scales while in the later it is enhanced. Then, we see a transition between the large scale spectrum to an enhanced (left) or suppressed (right) spectrum at smaller scales.}
\label{Plot: MG P(k)}
\end{figure*}

%%%%%%%%%%%%%%%%%%%%%%%%%%%%%%%%%%%%%%%%%%%%%%%%%%%%%%%%%%%%%%%%%%%%%
\section{The Matter Power Spectrum of Modified Gravity}
\label{Sec: The Linear Matter Power Spectrum of Modified Gravity}
%%%%%%%%%%%%%%%%%%%%%%%%%%%%%%%%%%%%%%%%%%%%%%%%%%%%%%%%%%%%%%%%%%%%%

As mentioned in Sec.~\ref{Sec: Introduction}, the existence of structure in our Universe follows from the agglomeration of matter due to gravitational instabilities. One of the key statistical correlation functions of this process is the matter power spectrum $P(k, z)$. It can be shown that, when non-linear effects are neglected, the matter power spectrum for a given redshift can be written as \cite{dodelson2020}
\begin{equation}
P(k) \equiv P_0 (k) T^2 (k)
\end{equation}
where $P_0(k)$ is an initial matter power spectrum and $T(k)$ is the matter transfer function. Now, we firstly comment on the form of $P(k)$ for $\Lambda$CDM, and then we discuss some of the possible alterations to this observable when modifications of gravity are considered.

In the standard case, the initial matter power spectrum is approximately given by 
\begin{equation}
\label{Eq: Initial Pk}
P_0  (k) \equiv A_0 k^{n_s}, \quad A_0 \equiv \frac{8 \pi^2}{25} \frac{\mathcal{A}_s}{\Omega_{m0}^2 H_0^4} k_p^{1-n_s} 
\end{equation}
where $\Omega_{m0}$ is the matter density parameter today, $H_0$ is the Hubble constant, $n_s$ and $\mathcal{A}_s$ are the spectral tilt and amplitude of primordial scalar perturbations, respectively, and $k_p$ is an arbitrary pivot scale \cite{dodelson2020}.\footnote{Note that $P_0(k)$ is not the primordial power spectrum which is given by $P_\mathcal{R}(k) \equiv \mathcal{A}_s (k/k_p)^{1 - n_s}$ \cite{Planck:2018jri}.} 
On the other hand, the matter transfer function takes into account the effects of all the interacting species, e.g. acoustic oscillations due to baryonic pressure, amplitude suppression due to the free streaming of massive neutrinos, transition from a radiation dominated epoch to a matter dominated epoch, and more. All these physical details are encompassed by the fitting functions presented in Ref.~\cite{Eisenstein:1997ik, Eisenstein:1997jh}, which achieve an accuracy of 1-2\% while needing around 30 different expressions in their full formulations. However, in Ref.~\cite{Orjuela-Quintana:2022nn}, two of the authors here shows that the GAs are able to find much more economical expressions while achieving the same accuracy range.

To the best of our knowledge, there are currently no similar analytical formulations available when considering alternative models. To find derive such a formula, our first step is to characterize the potential effects that MG theories introduce in the matter power spectrum compared to the concordance model. In the following, we undertake this characterization process. Firstly, a background history different to that of $\Lambda$CDM can change the growth factor, which in turn modifies the overall amplitude of $P(k)$. Secondly, a variation in the reduced matter density parameter $\omega_m$ implies a variation in the redshift and scale at the radiation-matter transition, denoted respectively as $z_\text{eq}$ and $k_\text{eq}$. This causes a shift in the position of the maximum of the matter power spectrum. Before continuing, we want to mention that it is possible to estimate $z_\text{eq}$ and $k_\text{eq}$ using the approximated expressions given in Ref.~\cite{Eisenstein:1997ik}. Nevertheless, as shown in the Appendix~\ref{App: The scale at radiation-matter equality}, the GAs are able to find expressions with an accuracy improved by two orders of magnitude while keeping simplicity. As a third effect, modified gravity can induce a clustering effective dark energy component when the structure formation process is occurring. This of course modifies the amplitude of $P(k)$, \cite{Sapone:2009mb, Sapone:2012nh, Sapone:2013wda}. Modifications at the largest scales due to ISW or normalization. Finally, there could be a variation in the amplitude and location of the acoustic peaks. However, these effects are principally controlled by the reduced baryon density parameter $\omega_b$ which is tightly constrained by observations \cite{Planck:2018vyg}. Here, we fix it to its default value in the popular Boltzmann solver \texttt{CLASS}; i.e. $\omega_b = 0.0223828$, and neglect the small oscillations in our formulation.

In Fig.~\ref{Plot: MG P(k)} we have 
plotted the matter power spectra for four different models varying some of their parameters in order to exemplify the physical effects introduced by modifications of gravity, as explained above. These plots were obtained using the branch \texttt{hi\_class} of \texttt{CLASS}. In the upper-left panel, we consider a $w$CDM model, with a constant speed of sound $c_s^2$, no anisotropic stress, and the equation of state of dark energy $w$ varying in the interval $[-1.2, -0.8]$. We can see that the overall amplitude of $P(k)$ is modified when $w$ varies. In the upper-right panel, we consider a $\Lambda$CDM model, varying the reduced CDM density parameter $\omega_c$ in the range $[0.11, 0.13]$. We note small shifts in scale of the maximum of $P(k)$. For the lower panels, we used one of the implemented models by default in \texttt{hi\_class}; namely, the \texttt{propto\_scale} model. In this  model, the effective field theory (EFT) parameters characterizing the cosmological dynamics of the whole Horndeski theory are taken as proportional to the scale factor. For more details about the EFT of Horndeski theories see Refs.~\cite{Bellini:2014fua, Bellini:2015xja}, and for the code \texttt{hi\_class} see Refs.~\cite{Zumalacarregui:2016pph, Bellini:2019syt}. In particular, for these plots we fixed the EFT parameters and vary only the proportionality constant of the braiding parameter $\hat{\alpha}_B$ (lower-left panel), and of the mass running parameter $\hat{\alpha}_M$ (lower-right), while the background evolution matches that of $\Lambda$CDM. As it can be seen, there is a clear distinction between the behavior of $P(k)$ at large scales and at small scales, with a marked transition occurring before the scale at equality. The matter power spectrum at the largest scales holds the form of the initial matter power spectrum proportional to $k^{n_s}$ while differences can be attributed to variations in the amplitude $A$. Then, at smaller scales, the spectrum can be enhanced (as in the lower-left panel) or suppressed (as in the lower-right panel) by modified gravity. In all the cases, we have depicted the matter power spectrum of $\Lambda$CDM to make notable the differences introduced by MG models.

In the next section, we propose a parameterized function for the matter power spectrum $P(k)$ that incorporates the aforementioned effects.

%%%%%%%%%%%%%%%%%%%%%%
\section{Results}
\label{Sec: Results}
%%%%%%%%%%%%%%%%%%%%%%

%%%%%%%%%%%%%%%%%%%%%%%%%%%%%%%%%%%%%%%%%%
\subsection{Parameterization of $P(k)$}
%%%%%%%%%%%%%%%%%%%%%%%%%%%%%%%%%%%%%%%%%%

Summarizing our discussion in the last section, we consider the following modified gravity effects on the $P(k)$:
\begin{enumerate}
\item Variations in the amplitude $A$ of the initial matter power spectrum $P_0(k)$ at large scales.
\item Overall enhancement or suppression of the amplitude.
\item Shift of the equality scale $k_\text{eq}$.
\end{enumerate}
In order to encompass these three effects in one single expression, we assume that the total matter power spectrum considers the contribution of two different spectra. At large scales, the most relevant contribution to $P(k)$ comes from the following expression 
\begin{equation}
\label{Eq: Plarge}
P_\text{large} (k; s, n_s) \equiv A(s) \left(\frac{k}{h/\text{Mpc}}\right)^{n_s} \left[ (\text{Mpc}/h)^3 \right],
\end{equation}
where  
\begin{equation} 
A(s) \equiv (1 + s) \frac{P_{\Lambda\text{CDM}}(k_i)}{k_i^{n_s} T_{\Lambda\text{CDM}}^2(k_i)},
\end{equation}\\
is a dimensional constant. The matter power spectrum $P_{\Lambda\text{CDM}}$ and the matter transfer function $T_{\Lambda\text{CDM}}$ for $\Lambda$CDM can be retrieved from e.g. \texttt{CLASS}, and thus $k_i$ corresponds to the smallest value of $k$ considered by the Boltzmann solver. Notice that this parameter is basically a modification of $A_0$ defined in Eq.~\eqref{Eq: Initial Pk}. However, we do not use $A_0$ directly in the definition of $A(s)$, i.e., we do not define this parameter as $A(s) \equiv A_0 (1 + s)$, since the expression for $A_0$ is an approximation only valid for sub-horizon modes \cite{dodelson2020}. The parameter $s$ takes on negatives values when the matter power spectrum is suppressed (as shown in the lower-left panel of Fig.~\ref{Plot: MG P(k)}), and positive values when the primordial matter power spectrum is enhanced (as depicted in the lower-right panel of Fig.~\ref{Plot: MG P(k)}) at large scales. The additional effects, such as an overall modification of the amplitude and the shift of the maximum, are incorporated into the matter power spectrum (referred to as $P_{\rm small}$), which primarily manifest at smaller scales. In the upper-right panel of Fig.~\ref{Plot: MG P(k)}, it is evident that a change in the parameter $w$ in the $w$CDM model can result in an overall enhancement or suppression of the amplitude of the matter power spectrum. Hence, we adopt this model as a reference to develop a formula for $P_{\rm small}$ utilizing the GAs. The accuracy of such a formula is measured by the following fitness function
\begin{equation}
\text{Acc}\% \equiv \left\langle \frac{|P_\text{CLASS} - P_\text{analytical}|}{P_\text{CLASS}} \right\rangle \times 100,
\label{Eq: Acc}
\end{equation}\\
where $P_\text{CLASS}$ is the matter power spectrum numerically computed using \texttt{CLASS}, and $P_\text{analytical}$ is the expression obtained from the GAs. In particular, we consider a constant sound speed given by $c_s^2 = 0.001$, fix the spectral tilt $n_s$ to the default value in \texttt{CLASS}, and vary $w$ in the range $[-1.2, -0.8]$. The GAs runs give the following expression\footnote{For details about the specific structure of the GAs, e.g., the genome arrangement and avoidance of overfitting problems, please refer to Ref.~\cite{Orjuela-Quintana:2022nn}.}
\begin{widetext}
\begin{align}
\label{Eq: Psmall}
P_\text{small} (k; \alpha, \omega_m) &= \frac{2.52395 \times 10^6 \left(\frac{k}{h/\text{Mpc}}\right)^{0.96452} \left(1 + 0.710636 \alpha^3 + 1.88019 \alpha^4 + 0.939217 \alpha^5 \right)}{\left(1 + 59.0998 \, x^{1.49177} + 4658.01 \, x^{4.02755} + 3170.79 \, x^{6.06} + 150.089 \, x^{7.28478}\right)^{1/2}} \, \left[ (\text{Mpc}/h)^3 \right], 
\end{align}
\end{widetext}
whose accuracy is $1.72\%$. In the last equation, we have defined the dimensionless variable $x \equiv k /(\omega_m - \omega_b) \, \text{Mpc}/h$, since $k$ is given in units $[h/\text{Mpc}]$, and we promoted $w$ to a general parameter $\alpha$ in order to avoid possible confusions with the notation. The parameter $\alpha$, in general, measures the overall modifications in the amplitude of $P(k)$ and it is not restricted to assume values in the range $[-1.2, -0.8]$. However, this parameter has the lower bound $\alpha > -1.68492$ since $P_\text{small}$ is positive for all $k$. It is important to emphasize that the $w$CDM model is used purely as a template in our study, as the variation of $w$ specifically generates the effect we aim to capture, which is an overall modification of the amplitude of the matter power spectrum.

Finally, we assume that the spectra in Eqs.~\eqref{Eq: Plarge} and \eqref{Eq: Psmall} are smoothly connected by a transition function given by
\begin{equation}
\label{Eq: transition function}
\sigma (k; k_0, \beta) = \left[ 1 + e^{-(\ln k - \ln k_0)/\beta} \right]^{-1}.
\end{equation}
where the transition is centered at $k_0$ and $\beta$ determines its width.\footnote{This transition function is not unique. Another common option for a ``smooth step function'' which would be valid is $\sigma(k; k_0, \beta) \equiv \left(1 + \tanh\left[(\ln k - \ln k_0)/\beta\right]\right)/2$.} It is noteworthy that the limits of this function are $\sigma \rightarrow 0$ when $k \ll k_0$, and $\sigma \rightarrow 1$ when $k \gg k_0$, and that we have defined $\sigma$ using logarithms since usually $P(k)$ is depicted in log-space. The transition between these limits occurs around $k \sim k_0$ with a smoothing level determined by $\beta$. Consequently, a small $\beta$ corresponds to a rapid transition, while a large $\beta$ results in a slower transition. Combining these considerations, the parameterized expression for $P(k)$ in the context of modified gravity is given by:
\begin{widetext}
\begin{align}
\label{Eq: P(k) of MG}
P_\text{MG} (k; \alpha, \beta, k_0, s, \omega_m, n_s) \equiv [1 - \sigma (k; k_0, \beta)] P_\text{large}(k; s, n_s) 
 + \sigma (k; k_0, \beta) P_\text{small}(k; \alpha, \omega_m). 
\end{align}
\end{widetext}
The new expression requires 6 parameters as input: the spectral tilt $n_s$, the shift $s$ at large scales, the overall enhancement (or suppression) parameter $\alpha$, the reduced matter density parameter $\omega_m$, the transition center $k_0$, and the transition width $\beta$ switching between $P_\text{large}$ and $P_\text{small}$. With the exception of $n_s$ and $\omega_m$, which are determined through observations, the remaining four parameters are free. We want to emphasize that the expression obtained through GAs was $P_\text{small}$ in Eq.~\eqref{Eq: Psmall}, while $P_\text{large}$ in Eq.~\eqref{Eq: Plarge} and $\sigma$ in Eq.~\eqref{Eq: transition function} are assumed and motivated by physical intuitions.

In the following section, we will evaluate the effectiveness of Eq.~\eqref{Eq: P(k) of MG} by estimating the best-fit values of parameters to accurately reproduce the numerical matter power spectra obtained from various MG theories using the \texttt{hi\_class} and \texttt{mgclass} codes.

%%%%%%%%%%%%%%%%%%%%%%%%%%
\subsection{Performance}
%%%%%%%%%%%%%%%%%%%%%%%%%%

\begin{center}
\begin{table*}
\begin{centering}
\begin{tblr}{X[1.5, c] X[2.5, c] X[c] X[c] X[c] X[c] X[c]}
\hline
\hline
Model & Parameter & $\alpha$ & $k_0 \times 10^{-3}$ & $\beta$ & $s$ & $\text{Acc}\%$ \\
\hline
\hline
\texttt{propto\_scale} & $\hat{\alpha}_B = 0.5$ & $0.68153$ & $1.16918$ & $0.42243$ & $-0.21975$ & $1.53\%$ \\
\hline
%%%%%%%%
\texttt{propto\_scale} & $\hat{\alpha}_M = 0.5$ & $0.36380$ & $0.73084$ & $0.45974$ & $0.53958$ & $1.62\%$  \\
\hline
%%%%%%%%
\texttt{propto\_scale} & $\hat{\alpha}_T = - 1$ & $0.53444$ & $0.40205$ & $0.39972$ & $-0.47066$ & $1.52\%$  \\
\hline
%%%%%%%%
\texttt{propto\_omega} & $\tilde{\alpha}_B = 0.625$ & $0.54866$ & $0.76693$ & $0.45681$ & $-0.07756$ & $1.46\%$  \\
\hline
%%%%%%%%
\texttt{propto\_omega} & $\tilde{\alpha}_M = 3.0$ & $0.59006$ & $0.39858$ & $0.46187$ & $1.57924$ & $1.48\%$  \\
\hline
%%%%%%%%
\texttt{propto\_omega} & $\tilde{\alpha}_T = -0.25$ & $0.53215$ & $1.16323$ & $0.42635$ & $0.02292$ & $1.47\%$  \\
\hline
%%%%%%%%
\texttt{plk\_late} & $E_{11} = 0.5, \, E_{22} = 0.5$ & $0.54762$ & $0.76681$ & $0.49599$ & $-0.62532$ & $1.43\%$ \\
%%%%%%%%
 & $E_{11} = -0.5, \, E_{22} = -0.5$ & $0.42644$ & $0.28607$ & $0.51071$ & $2.99343$ & $1.51\%$ \\
\hline
%%%%%%%%
HDES & $\tilde{J}_c = 0.005$ & $0.56331$ & $2.24347$ & $0.443931$ & $0.05417$ & $1.39\%$  \\
\hline
\end{tblr}
\par\end{centering}
\caption{Best-fit parameters $\alpha$, $k_0$, $\beta$, and $s$ for several models in the codes \texttt{hi\_class} and \texttt{mgclass}. Just one value of the parameters of each model is shown to keep our presentation simple. We can see that the mean accuracy in each example is around 1-2\%.}
\label{Tab: Models}
\end{table*}
\par\end{center}

Here, we numerically compute matter power spectra using the codes \texttt{hi\_class} and \texttt{mgclass} considering several models. Subsequently, we find the best-fit values of the parameters $\alpha$, $k_0$, $\beta$, and $s$ by minimizing the fitness function in Eq.~\eqref{Eq: Acc}. This procedure will yield to the best match of $P_\text{MG}(k)$ in Eq.~\eqref{Eq: P(k) of MG} with respect to the data from \texttt{CLASS}. 

There are several models already implemented in \texttt{hi\_class}, and we make use of two specific ones: \texttt{proto\_scale} and \texttt{propto\_omega}. In the \texttt{propto\_scale} model, the EFT parameters of the Horndeski theory are proportional to the scale factor $a$, i.e. to $\alpha_i \equiv \hat{\alpha}_i a$, where $i = K, B, M, T$ denotes the kineticity, braiding, mass running and tensor speed excess parameters, respectively. On the other hand, in the \texttt{propto\_omega} model, the EFT parameters are proportional to the dark energy density parameter $\Omega_\text{DE}$, i.e. $\alpha_i \equiv \tilde{\alpha}_i \Omega_\text{DE}$. We also add to \texttt{hi\_class} the designer Horndeski model (HDES) introduced in Ref.~\cite{Arjona:2019rfn}. The HDES model has two parameters; namely, $\tilde{J}_c$ and $n$. We fix $n = 1$, and consider a small value for $\tilde{J}_c$. In all these models, it is assumed that the background evolution exactly matches that of $\Lambda$CDM. In \texttt{mgclass}, we also use an already implemented model called \texttt{plk\_late}, in which the slip parameters are parameterized in terms of $\Omega_\text{DE}$ (see Ref.~\cite{Sakr:2021ylx}). This model in particular has been used by observational collaborations such as Planck \cite{Planck:2015bue} and DES \cite{Abbott:2018xao}. 

\begin{figure}[t]
\centering
\includegraphics[width=0.48\textwidth]{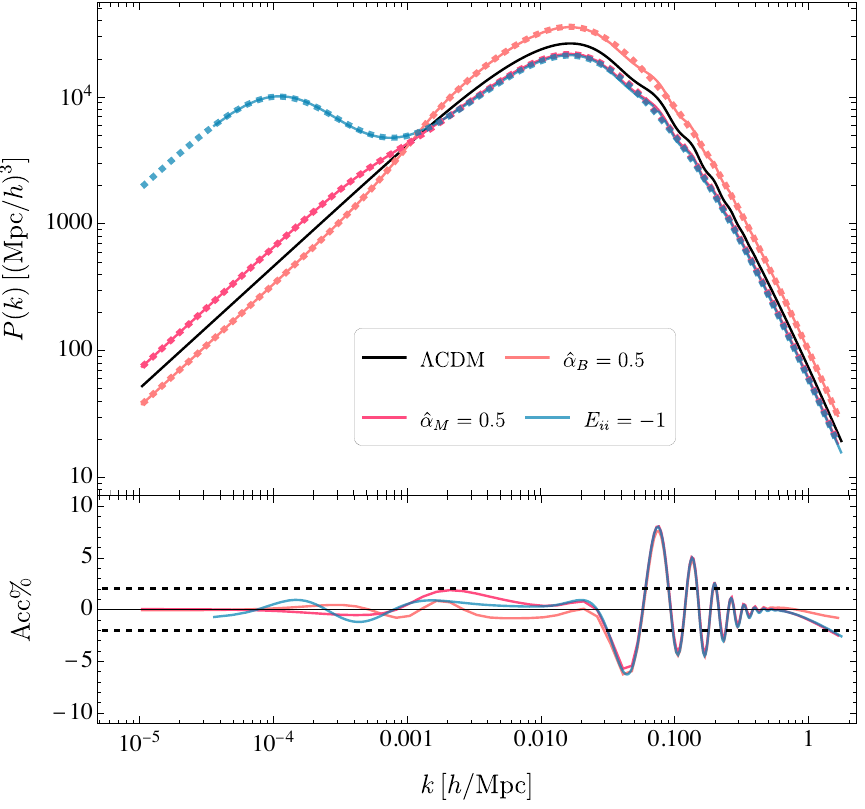}
\caption{Three examples showing the fitting of our formula in Eq.~\eqref{Eq: P(k) of MG} (dots) with respect to numerical results (solid lines). The matter power spectrum of $\Lambda$CDM (black line) is plotted to highlight the effects introduced by MG theories. The subplot shows the accuracy of the formulae as a function of $k$. We can see that most of points are in the 2\% region defined by the black dotted lines.}
\label{Plot: Pk Fits}
\end{figure}

In Table~\ref{Tab: Models}, we show the best-fit values of the parameter set $\{\alpha, k_0, \beta, s\}$ along with the associated mean accuracy, achieved by selecting a specific parameter value from each model. It is important to mention that we have explored multiple parameter values for each model. However, for the sake of simplicity in the presentation, we report only one value. The details of the analysis and the parameter exploration for each model can be found in the provided codes (see the Numerical Codes statement). Table~\ref{Tab: Models} shows that the mean accuracy for the considered models is approximately $1$-$2\%$.\footnote{We want to mention that since we fix $n_s$ to find $P_\text{small}$ using the GAs, this could yield to errors when different values of $n_s$ are considered. Although, the effect of $n_s$ on $P(k)$ is small, we verified that the mean accuracy of our formula $P_\text{MG}$ is still around $1\%$ when $n_s$ is varied in the interval $[0.96, 0.97]$ which is $1\sigma$ away from the value measured by Planck \cite{Planck:2018vyg}.} In Fig.~\ref{Plot: Pk Fits}, we plot three examples of the matter power spectra from MG theories to visually show the excellent fit provided by our formula with respect to numerical data. In this plot, we use the models \texttt{propto\_scale} of \texttt{hi\_class} varying the braiding constant $\hat{\alpha}_B$ and the mass running constant $\hat{\alpha}_M$. From \texttt{mgclass}, we used the \texttt{plk\_late} model assuming the parameters $E_{11} = E_{22} = -1.0$. We also plot the $\Lambda$CDM matter power spectrum to highlight the modifications introduced by these alternative theories. In the subplot, we depict the accuracy as a function of $k$, which can be measure as
\begin{equation}
\label{Eq: Acck}
\text{Acc}\% (k) \equiv \left(\frac{P_\text{CLASS} - P_\text{analytical}}{P_\text{CLASS}} \right)  \times 100,
\end{equation}
where no mean value is taken, as in Eq.~\eqref{Eq: Acc}. This shows that most of values are in the 2\% region defined by the black dotted lines. Remarkably, our formula is able to capture the substantial deviation at large scales observed in the spectrum of the \texttt{plk\_late} model (depicted in blue) and the significant enhancement around $k_\text{eq}$ in the \texttt{propto\_scale} braiding model (depicted in pink). It is also noteworthy that although the expression of $P_\text{small}$ in Eq.~\eqref{Eq: Psmall} does not consider the wiggles introduced by BAOs in the matter power spectrum, the error of neglecting this oscillatory behavior is around 5\% in the most inaccurate cases. The amplitude of the acoustic peaks depends mainly on the amount of baryonic matter, which translate to the parameter $\omega_b$. However, observations indicate that $\omega_b \approx 0.022$ implying small peaks. Therefore, a line passing through these oscillations seems to be a good approximation.\footnote{However, the formula for $P_\text{MG}(k)$ should not be used when an accurate description of these wiggles is required. The introduction of these oscillations in $P_\text{MG}(k)$ is a further improvement which we leave for a future work.}

%%%%%%%%%%%%%%%%%%%%%%%%%
\section{Conclusions} 
\label{Sec: Conclusions}
%%%%%%%%%%%%%%%%%%%%%%%%%

Accurate fitting formulations for the matter transfer function allow to compute the matter power spectrum within the context of the $\Lambda$CDM model \cite{Bardeen:1985tr, Eisenstein:1997ik, Eisenstein:1997jh, Orjuela-Quintana:2022nn}. However, there is no any formulation available when modifications of gravity are considered. In this work, we have put forward a parametric function for $P(k)$ considering several effects introduced by modified gravity theories. The new formula for $P_\text{MG}(k)$ was presented in Eq.~\eqref{Eq: P(k) of MG} and it is composed by two main parts. A matter power spectrum $P_\text{large}$ encompassing modifications at large scales, and the spectrum $P_\text{small}$ which manifest at smaller scales. The former is proportional to $k^{n_s}$ with amplitude $A(s) \approx A_0 (1 + s)$, where $s$ is a parameter measuring the shift of the initial amplitude $A_0$ which in the case of $\Lambda$CDM is approximately given in Eq.~\eqref{Eq: Initial Pk}. The other spectrum; namely $P_\text{small}$, was obtained through the machine learning technique of the genetic algorithms using a $w$CDM model as template. The expression for $P_\text{small}$ in Eq.~\eqref{Eq: Psmall} involves two parameters: a parameter $\alpha$ measuring overall modifications (enhancement or suppression) in the amplitude of the spectrum, and $\omega_m$ which controls the value of the scale at equality $k_\text{eq}$. These two spectra; $P_\text{large}$ and $P_\text{small}$, are smoothly connected by the transition function $\sigma$ defined in Eq.~\eqref{Eq: transition function}. This function marks the transition center at the scale $k_0$ with a smoothness measured by the parameter $\beta$. Therefore, our formulation requires 6 input parameters: the spectral tilt $n_s$, the shift $s$ of $A_0$ at large scales, the enhancement (suppression) factor $\alpha$ measuring modifications in the amplitude of $P_\text{small}(k)$, the reduced matter density parameter $\omega_m$, and two parameters, $k_0$ and $\beta$, that determine the center and width of the transition region between the matter power spectra $P_\text{large}$ and $P_\text{small}$ [see Eq.~\eqref{Eq: P(k) of MG}].

The performance of our formulation was compared against numerical results obtained from the codes \texttt{hi\_class} and \texttt{mgclass} for several models. The relatively simple expression in Eq.~\eqref{Eq: P(k) of MG} showed a mean accuracy of about $1$-$2\%$ (see Table~\ref{Tab: Models}). In Fig.~\ref{Plot: Pk Fits}, we showed three examples of the fitting of our formula with respect to some MG models. In the subplot, the accuracy of the formulae as a function of $k$ using Eq.~\eqref{Eq: Acck} was measured. This showed that most of values predicted by our formula lies in the $2\%$ region. Furthermore, although our expression does not consider baryonic acoustic oscillations, neglecting these wiggles introduces an error not greater than 5\%. Given that ongoing surveys, like Euclid, expect to reach an observational accuracy of around 1\% for some cosmological parameters which are obtained using the matter power spectrum, and that they already use the Eisenstein-Hu formula in their analysis~\cite{Euclid:2019clj}, our formula could extend these analysis to model beyond $\Lambda$CDM without loosing accuracy. Therefore, the function $P_\text{MG} (k)$ in Eq.~\eqref{Eq: P(k) of MG} represents a compelling fitting formula for the matter power spectrum of alternative models to $\Lambda$CDM.

While this solution does not possess the baryonic wiggles necessary for inferring cosmological parameters from the BAO signal, it can still be used for parameter inference. To illustrate this, let us consider the approach used to forecast cosmological parameters for the Euclid survey \cite{Euclid:2019clj}. The observed galaxy power spectrum relies on the de-wiggled matter power spectrum, which is the sum of two components: the linear matter power spectrum obtained from a Boltzmann code and the linear matter power spectrum without wiggles but exhibiting the same broad band as the linear matter power spectrum. Both components are modulated by a function that controls the non-linear damping of the BAO signal in all directions. The no-wiggles power spectrum  is often computed using the a fitting formula (for instance, for the $\Lambda$CDM models, the EH). It is clear that, when exploring cosmological models beyond the standard framework, it becomes crucial to appropriately parameterize the no-wiggles matter power spectrum in order to accurately capture the amplitude and shape of the observed galaxy power spectrum.

%%%%%%%%%%%%%%%%%%%%%%%%%%%%%
\section*{Acknowledgements}
%%%%%%%%%%%%%%%%%%%%%%%%%%%%%

BOQ acknowledges support from Patrimonio Aut\'onomo - Fondo Nacional de Financiamiento para la Ciencia, la Tecnolog\'ia y la Innovaci\'on Francisco Jos\'e de Caldas (MINCIENCIAS - COLOMBIA) Grant No. 110685269447 RC-80740-465-2020, projects 69723 and 69553. SN acknowledges support from the research project PID2021-123012NB-C43, and by the Spanish Research Agency (Agencia Estatal de Investigaci\'on) through the Grant IFT Centro de Excelencia Severo Ochoa No CEX2020-001007-S, funded by MCIN/AEI/10.13039/501100011033. DS acknowledges financial support from the Fondecyt Regular project number 1200171.

%%%%%%%%%%%%%%%%%%%%%%%%%%%%%
\section*{Numerical Codes}
\label{Numerical Codes}
%%%%%%%%%%%%%%%%%%%%%%%%%%%%%

The genetic algorithm code used here can be found in the GitHub repository \href{https://github.com/BayronO/Pk-in-Modified-Gravity}{https://github.com/BayronO/Pk-in-Modified-Gravity} of BOQ. This code is based on the GA code by SN which can be found at \href{https://github.com/snesseris/Genetic-Algorithms}{https://github.com/snesseris/Genetic-Algorithms}.

%%%%%%%%%%%
\appendix
%%%%%%%%%%%

%%%%%%%%%%%%%%%%%%%%%%%%%%%%%%%%%%%%%%%%%%%%%%%%%%%%%%
\section{Genetic Algorithm Expressions for $k_\text{eq}$ and $z_\text{eq}$}
\label{App: The scale at radiation-matter equality}
%%%%%%%%%%%%%%%%%%%%%%%%%%%%%%%%%%%%%%%%%%%%%%%%%%%%%%

Variations in the parameter $\omega_m$ implies in general a shift of the radiation-matter transition. In Ref.~\cite{Eisenstein:1997ik}, the scale and redshift at this transition were given by, 
\begin{align}
z_\text{eq}(\omega_m) &= 2.5 \times 10^4 \Theta^{-4}_{2.7} \omega_m, \\
k_\text{eq}(\omega_m) &= 7.46 \times 10^{-2} \Theta^{-2}_{2.7} \omega_m,
\end{align}
where $\Theta_{2.7}$ defines the CMB temperature through the relation $T_\text{CMB} \equiv 2.7 \Theta_{2.7} \, \text{K}$. Taking $T_\text{CMB} = 2.7255 \, \text{K}$ we have the approximated expressions:
\begin{align}
k_\text{eq} &= 0.0732106 \omega_m, \\
z_\text{eq} &= 24077.4 \omega_m.
\end{align}

In the case of $\Lambda$CDM, we used \texttt{CLASS} to get data for $k_\text{eq}$ and $z_\text{eq}$ varying $\omega_m$ in the range $[0.13, 0.15]$, which is around $10\sigma$  from the Planck best-fit. Comparing these datasets with the above formulae, we find that the accuracy of $k_\text{eq}$ is around $0.37\%$, while for $z_\text{eq}$ we get $0.74\%$. Although these formulae are accurate, the GA was able to find even better analytical formulations. The GAs ended up with the following expressions:
\begin{align}
k_\text{eq}(\omega_m) &= 0.07222 \omega_m + 0.00151 \omega_m^{1.3775}, \\
z_\text{eq}(\omega_m) &= 23896.7 \omega_m + 11.6353 \omega_m^{1.41595}, 
\end{align}
The accuracy of $k_\text{eq}$ is 0.023\%, while for $z_\text{eq}$ the GAs achieve an accuracy of 0.0025\%, representing an improvement of 1 and 2 orders of magnitude respectively.

\bibliographystyle{utcaps} 
\bibliography{Bibli.bib}

\end{document}